\documentclass[aip,jcp,reprint,floatfix,twocolumn]{revtex4-1}

\usepackage{times}
\usepackage{graphicx}
\usepackage{amsmath}
\usepackage{amssymb}
\usepackage{color}
\usepackage{url}

\newcommand{\ham}{\hat{\mathcal{H}}}
\newcommand{\bra}[1]{\ensuremath{\left<{#1}\right|}}
\newcommand{\ket}[1]{\ensuremath{\left\vert{#1}\right>}}
\newcommand{\quotes}[1]{``#1''}

\begin{document}
\raggedbottom


\title{Perturbatively selected configuration-interaction wave functions for efficient geometry optimization in quantum Monte Carlo}

\author{Monika Dash}
\affiliation{MESA+ Institute for Nanotechnology, University of Twente, P.O. Box 217, 7500 AE Enschede, The Netherlands}
\author{Saverio Moroni}
\email{moroni@democritos.it}
\affiliation{CNR-IOM DEMOCRITOS, Istituto Officina dei Materiali, and SISSA Scuola Internazionale Superiore di Studi Avanzati, Via Bonomea 265, I-34136 Trieste, Italy}
\author{Anthony Scemama}
\email{scemama@irsamc.ups-tlse.fr}
\affiliation{Laboratoire de Chimie et Physique Quantiques, Universit\'e de Toulouse, CNRS, UPS, France}
\author{Claudia Filippi}
\email{c.filippi@utwente.nl}
\affiliation{MESA+ Institute for Nanotechnology, University of Twente, P.O. Box 217, 7500 AE Enschede, The Netherlands}

\begin{abstract}

We investigate the performance of a class of compact and systematically
improvable Jastrow-Slater wave functions for the efficient and accurate
computation of structural properties, where the determinantal component is
expanded with a perturbatively selected configuration interaction scheme
(CIPSI).
We concurrently  optimize the molecular
ground-state geometry and full wave function -- Jastrow factor, orbitals, and
configuration interaction coefficients-- in variational Monte Carlo (VMC) for the
prototypical case of 1,3-trans-butadiene, a small yet theoretically challenging
$\pi$-conjugated system.  We
find that the CIPSI selection outperforms the conventional scheme of
correlating orbitals within active spaces chosen by chemical intuition: it
gives significantly better variational and diffusion Monte Carlo energies for
all but the smallest expansions, and much smoother convergence of the geometry
with the number of determinants.  In particular, the optimal bond lengths and
bond-length alternation of butadiene are converged to better than one m\AA\
with just a few thousand determinants, to values very close to the
corresponding CCSD(T) results.  The combination of CIPSI expansion and VMC
optimization represents an affordable tool for the determination of accurate
ground-state geometries in quantum Monte Carlo.    
\end{abstract}

\maketitle

\section{Introduction}
\label{sec:intro}

Quantum Monte Carlo methods are a class of {\it ab initio} approaches which
solve the interacting Schr\"odinger equation stochastically. The most widely
used variants of QMC are the variational (VMC) and diffusion Monte Carlo (DMC).
Thanks to their favorable scaling with the number of particles and the ease of
parallelization, they have often been employed to benchmark electronic
properties, in particular, total energies of relatively large molecules as well
as solids. Recently, it has been shown~\cite{filippi2016,assaraf2017} that it
is possible to compute derivatives of the energy at the same computational cost
per Monte Carlo step as evaluating the energy itself, also when employing large
determinantal expansions in the commonly used Jastrow-Slater QMC wave
functions. Consequently, one can compute all derivatives necessary for the
optimization of the structure of a system very efficiently and, simultaneously,
of the variational parameters in the wave function, as was demonstrated on the
structural optimization of short polyenes with expansions comprising over
200,000 determinants~\cite{assaraf2017}.  These developments 
also allow us to thoroughly explore
the sensitivity of QMC calculations to the choice of the Slater expansion,
namely, the set of orbitals which one must correlate and the truncation of the
active space. It was shown that an instructed guess of
the orbitals based on chemical intuition can lead to significant variations in
VMC energies and structures of a molecule as small as butadiene even when
employing large expansions in the presence of a Jastrow correlation
factor~\cite{assaraf2017}.

To overcome the limitations of an {\it a priori} approach in the choice of the
orbital set and achieve a compact description of the determinantal component in
QMC, a promising alternative is to employ a selected CI algorithm such as the
CIPSI (configuration interaction using a perturbative selection done
iteratively) method. CIPSI was originally introduced by Huron {\it et al.}\ in
1973~\cite{huron1973} and continually
tested~\cite{gouyet1976,povill1992,trinquier1981,castex1981,pelissier1981,nebot1981,cimiraglia1987},
improved~\cite{evangelisti1983,cimiraglia1985,cimiraglia1985ab,cimiraglia1987},
and assessed in comparison to full-CI (FCI)
expansions~\cite{illas1991,harrison1991,cimiraglia1987} in combination with a
variety of orbital descriptions~\cite{illas1988}. In the last few years, there
has been renewed interest in the development of selected CI
approaches~\cite{ohtsuka2017,schriber2017,schriber2016,holmes2016,sharma2017}
to accurately calculate the ground- and excited-state energies of small
molecules, establishing the competence of these approaches for benchmarking
applications.  
The use of these expansions has however only been marginally explored within
the QMC
framework~\cite{giner2013,scemama2014,giner2015,giner2016,caffarel2016comm,scemama2018,caffarel2016comm},
where it was shown to yield very good DMC energies at the price of employing
large CIPSI wave functions.

Here, we complement the perturbatively selected CIPSI determinants with a
Jastrow factor and VMC optimization of the full wave function (Jastrow
parameters, orbitals and CI coefficients), and investigate 
the ability of
the resultant Jastrow-CIPSI wave functions to obtain accurate molecular
geometries in VMC, as well as corresponding VMC and DMC energies, with
relatively compact expansions. 
We focus on the butadiene molecule, where an accurate estimation of the bond
length alternation (BLA) is theoretically quite challenging, demanding a proper
description of correlation effects, both static and dynamic.  As mentioned
above, significant variations in energies and structural parameters were
observed in VMC when correlating different sets of $\sigma$ and $\pi$ orbitals
in a fully optimized Jastrow-Slater wave function~\cite{assaraf2017}: with a
large expansion comprising over 45,000 determinants in a restricted active
space, good agreement with coupled cluster with singles, doubles, and
perturbative triples in the complete basis set limit (CCSD(T)/CBS) was obtained
in the bond lengths.  This number of determinants is however surprisingly large
for the description of the ground state of such a small system. 
Indeed, we find here that a determinantal description with CIPSI yields a much
smoother
convergence in the structural parameters and QMC energies when the size of the
expansion is systematically incremented. Correspondingly, very accurate
values for these physical properties are already obtained with only a few
thousand determinants.

To construct the CIPSI component of the QMC trial wave
functions, we follow two different schemes: one where we
systematically \quotes{expand} the wave function by adding important
determinants at every step, and the other where we first generate an extremely
large wave function and subsequently \quotes{truncate} it to obtain requisite
sizes.
For a fixed size of the CI expansion, the \quotes{truncation} scheme is expected
to be more accurate than the \quotes{expansion} one since the overlap of the
wave function with the FCI wave function is larger. The difference can be
important in systems where the CI coefficients change dramatically with the
number of determinants in the wave function\cite{caffarel2014}, but
generation of the large wave function to initiate \quotes{truncation} might 
not be entirely
feasible for larger systems where the \quotes{expansion} scheme would therefore
represent a computationally less expensive route. For small systems such as butadiene, 
however, we can employ both strategies and, consequently, draw an
assessment of their relative performance.  Here, we find that the truncation scheme
provides a faster and smoother convergence of the bond lengths and VMC
energies of butadiene. We note that we are looking at extremely small variations in the
structural parameters (well below m\AA) when comparing and establishing the convergence. 
Finally, while we primarily employ multi-configurational self consistent field (MCSCF) 
canonical orbitals in the CIPSI algorithm and as starting orbitals in QMC,
we additionally test the use of natural orbitals. Rather peculiarly, expansions with natural orbitals
commensurate in size with those with canonical orbitals consistently converge to slightly
higher VMC energies.  
   
The paper is organized as follows. In section~\ref{sec:methods}, we
describe the CIPSI algorithm and the functional form of the Jastrow-Slater
wave functions and, in Section~\ref{sec:comput}, we report the computational details.
The numerical results obtained for the structural optimization of butadiene
and a comparison with our previous Jastrow-CAS results~\cite{assaraf2017}
are given in Section~\ref{sec:results}.

\section{Methods}
\label{sec:methods}

CIPSI is an iterative CI selection algorithm that allows us to perturbatively
select determinants from the FCI space. Starting with an initial
reference wave function, additional determinants are added to the expansion
based on their effective second-order energy contribution which is required to be
greater than a fixed or an iteratively modifiable threshold. Selection
iterations can be performed until a target number of determinants is reached
or until some other selection criterion is met~\cite{angeli1997}. In the
process, the quality of the wave function is systematically improved
and converges towards the FCI solution. 

A step-by-step description of the CIPSI algorithm is provided in
Refs.~\onlinecite{giner2016,caffarel2016,giner2015,giner2013}. The scheme is briefly
iterated here. The reference wave function is composed of a linear
combination of Slater determinants, $D_{i}$, spanning a space, $S$:
\begin{equation}
{\psi_{\rm CIPSI}} = \sum_{D_i\in \emph{S}} c_{i}{D_{i}}\,.
\end{equation}  
Typically, this reference wave function is initially the single determinant with the
lowest energy, which is also the case here. 
At every iteration,
the many-body Hamiltonian, ${\ham}$, is diagonalized in the reference space $S$ to obtain the
variationally minimized energy $E^{\rm ref}$ and the corresponding $c_{i}$ coefficients.
Then, all the determinants $D_j$ outside of this reference space that are connected to $S$
by $\ham$ are generated, and their individual contributions to the energy are
estimated with the Epstein-Nesbet~\cite{epstein1926,nesbet1955} perturbation theory,
\begin{equation}
\delta E_{j}^{(2)} = \frac{|\bra{D_j}\ham\ket{\psi_{\rm CIPSI}}|^2}{E^{\rm ref} - \bra{D_j}\ham\ket{D_j}}\,.
\end{equation}
If $|\delta E_j^{(2)}|$ is greater than a given threshold, determinant $D_j$ is selected 
for the next iteration.
Summing all these contributions gives $E^{\rm PT2}$, namely, the second-order perturbative energy correction
to $E^{\rm ref}$, and the CIPSI energy of the current iteration is given by
\begin{equation}
E^{\rm CIPSI} = E^{\rm ref}+E^{\rm PT2}.
\label{Ecipsi}
\end{equation}
Finally, all the selected determinants are added to the reference space $S$ for the next iteration.

For the generation of different lengths of the CIPSI component of the QMC
wave function, one could either follow (a) an \quotes{expansion} scheme where we repeat
the above process as many times as necessary, thereby systematically increasing
the size of the wave function, or (b) a \quotes{truncation} scheme after generating a
very large CIPSI wave function, typically to a point where the CIPSI
energy (Eq.~\ref{Ecipsi}) is reasonably converged. The large wave function created for 
the second scheme can contain millions of determinants and one needs to reorder them in
decreasing order of the absolute value of their CI coefficients and then
truncate the expansion, keeping their coefficients the same as in the large CIPSI
wave function. This strategy is said to be a better starting point for VMC
optimization as the determinant coefficients are from a near FCI
calculation~\cite{giner2016}. 
We employ the expansion scheme while keeping the wave function an eigenstate of
$\textbf{S}^{2}$ while no such constraint is used when following the truncation
scheme. Hence, the comparison of the two schemes is not exactly one to one. We use
complete active space SCF (CASSCF) orbitals for the majority of these calculations, however, (c) the use of
natural orbitals obtained from a large CIPSI calculation is another aspect of
our investigation.

After generating the CIPSI expansion, we introduce a positive Jastrow correlation
factor ${\cal J}$ and construct the Jastrow-CIPSI
wave function as
\begin{equation}
\psi = {\cal J}{\psi_{\rm CIPSI}} = {\cal J} \sum_{i=0}^{N_{\rm det}} c_{i}D_{i}\,,
\end{equation}
where ${\cal J}$ explicitly describes electron-electron and electron-nucleus (two-body) 
and electron-electron-nucleus (three-body) correlations~\cite{filippi1996a} while imposing the
electron-electron cusp conditions. $N_{\rm det}$ represents the number of determinants in the CIPSI
wave function spanning the space \emph{S}.


\section{Computational Details}
\label{sec:comput}

The QMC calculations are carried out with the program package CHAMP~\cite{umfil}. 
We employ scalar-relativistic energy-consistent Hartree-Fock 
pseudopotentials and the cc-pVTZ Gaussian basis set specifically constructed 
for our pseudopotentials~\cite{burkatzki2007,BFD_H2013}.  
In particular, we perform all calculations with the cc-pVTZ basis set and test
the convergence of the results with the cc-pVQZ basis set.
To generate the Slater component of the QMC
wave functions, CIPSI calculations are performed in Quantum
Package~\cite{scemama2015} using canonical orbitals obtained from a
CAS(10,10) MCSCF calculation for the ground state of butadiene carried out with
the program GAMESS(US)~\cite{schmidt1993,gordon2005}. 

All parameters (Jastrow, orbital, and CI coefficients) are variationally optimized 
in VMC using the stochastic reconfiguration method~\cite{neuscamman2012} in a conjugate gradient
implementation~\cite{sorella2007}. Most calculations are performed with a two-body
Jastrow factor and the impact of the electron-electron-nucleus terms on the energy is
tested for a few cases.  Exploiting the low-numerical-scaling 
computation of energy and wave function derivatives~\cite{assaraf2017}, the ground-state
geometry of butadiene is simultaneously optimized with the wave function
following the path of steepest descent and an appropriate rescaling of the
inter-atomic forces. To avoid spikes in forces, an improved estimator of the 
inter-atomic forces
is used in all calculations, which is obtained by sampling the square of a
modified wave function close to nodes~\cite{attaccalite2008}.
In the DMC calculations, we treat the pseudopotentials beyond the locality 
approximation~\cite{casula2006a} and use an imaginary time step of 0.015 a.u. As 
shown in the Supplementary Information (SI), this time step yields DMC energies
converged to better than 0.1 mHartree for a simple Jastrow-CIPSI wave function 
with only two determinants and is therefore appropriate for all
wave functions of higher quality considered in this work.

We employ Gaussian09~\cite{gaussian09} to perform CCSD(T) geometry 
optimization in combination with our pseudopotentials and in all-electron calculations 
in the frozen core (FC) approximation with the cc-pVXZ (X = D, T, Q, and 5) and aug-cc-pVXZ (X = D, T, and Q) 
basis sets.  For CCSD(T) geometry optimization without the FC approximation, we use the
the PSI4 code~\cite{parrish2017} with the cc-pCVXZ, cc-pwCVXZ, and corresponding
augmented (X = D, T, and Q) basis sets. The results of these optimizations and their 
extrapolations to the CBS limit are detailed in the SI. 

\section{Results and Discussion}
\label{sec:results}

We investigate the merits of the choice of a CIPSI expansion to describe the
Slater component of a QMC wave function when optimizing the ground-state
geometry of butadiene over the conventional CAS description.
We start all 
structural optimizations with the Jastrow-CIPSI wave functions from the same 
initial MP2/cc-pVQZ optimized geometry and, as described above, variationally optimize the Jastrow 
parameters, CI coefficients, and orbitals simultaneously with the geometry in VMC.
Post convergence of the VMC energy and stabilization of the bond lengths, we perform 
40 additional iterations and average these geometries. A final VMC and DMC
energy calculation is done on this average geometry with the
wave function obtained in the last iteration. As detailed in the SI, reoptimizing
the wave function on the average geometry leads to equivalent energies within the
statistical error.

\begin{figure}[b!]
\includegraphics[width=1.0\columnwidth]{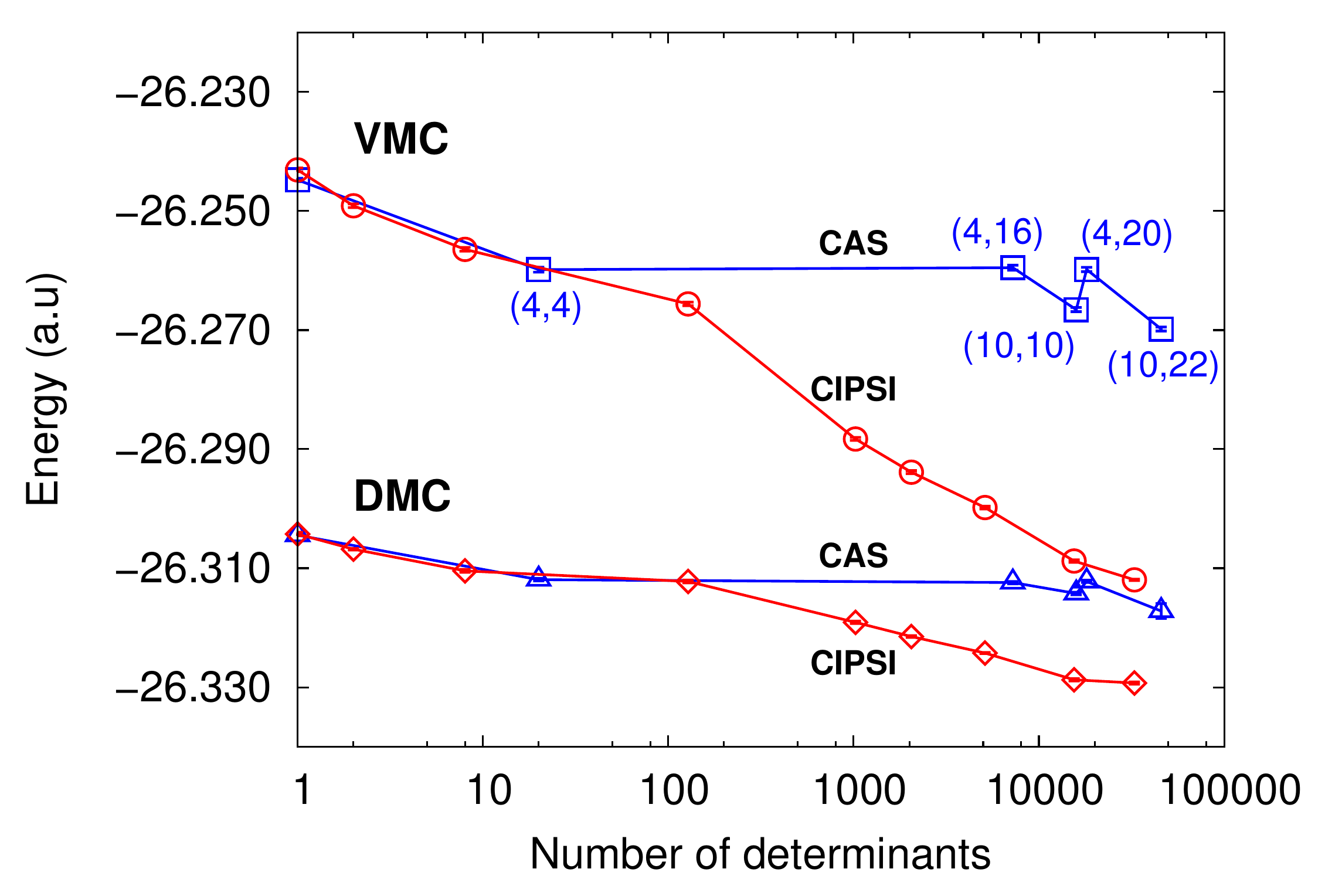}
\caption{Convergence of the VMC and DMC energies on the optimal geometry with the number of determinants in the CIPSI expansion (red). Previous results~\cite{assaraf2017} obtained with Jastrow-CAS wave functions are also presented (blue).}
\label{fig:energy1}
\end{figure}

\begin{figure}[!h]
\includegraphics[width=1.0\columnwidth]{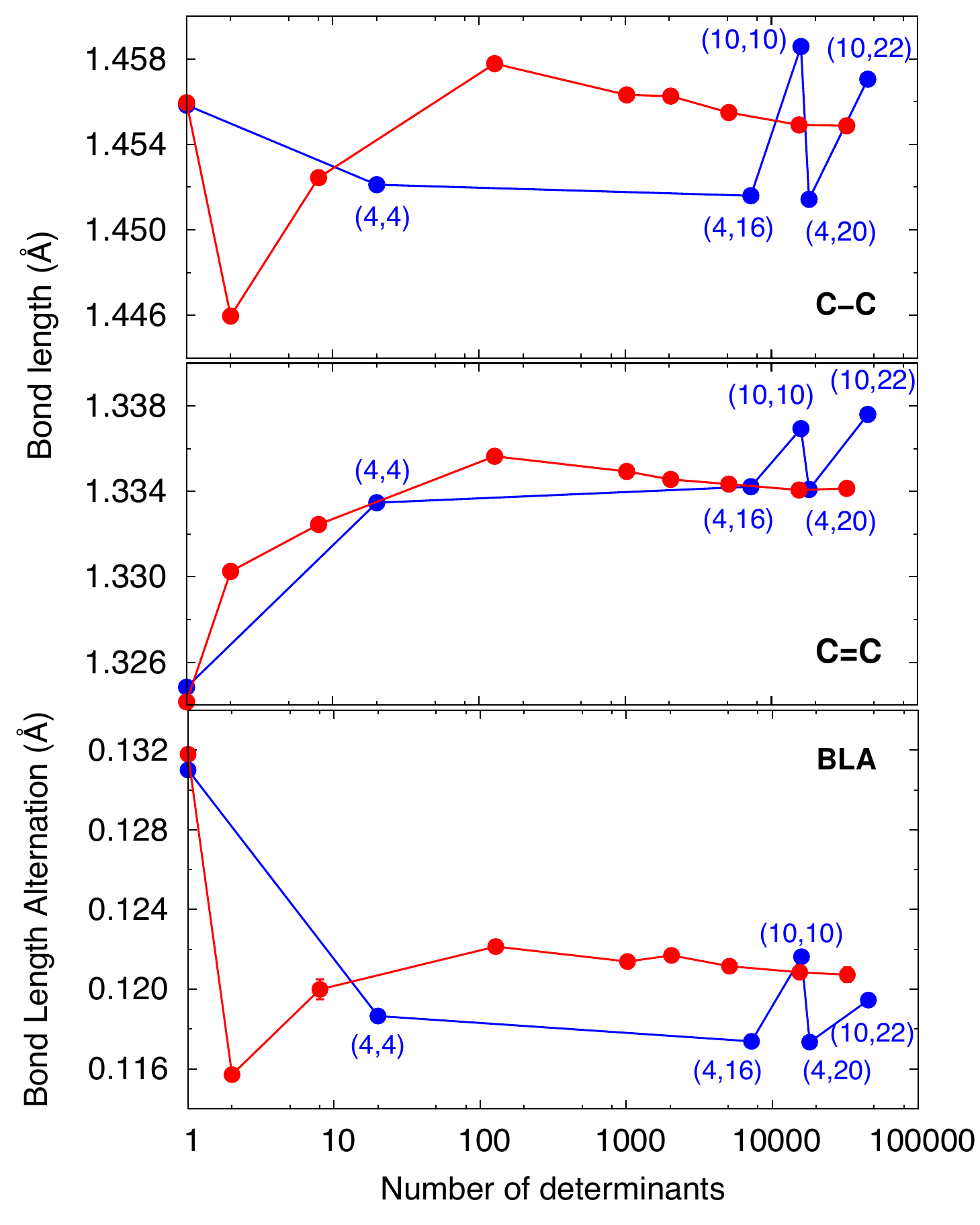}
\caption{Convergence of the single bond (C-C), double bond (C=C), and bond length alternation (BLA) (\AA) with the number of determinants in the CIPSI wave function (red). Previous results~\cite{assaraf2017} obtained with a Jastrow-CAS wave function are presented in blue.}
\label{fig:geomparams1}                                                  
\end{figure}

We primarily focus our discussion on the use of the \quotes{truncation} scheme for the CIPSI-determinant 
selection since, unlike the expansion scheme, no added selection criterion has been used. An initial CIPSI
wave function is constructed including as many as 1.17 million determinants and then truncated 
to generate a set of determinantal expansions of increasing size.
The results are summarized in Table~\ref{tab:geom-buta-vmc-dmc-pvtz} and illustrated
in Figs.~\ref{fig:energy1} and~\ref{fig:geomparams1}, where the final VMC and DMC energies, 
and corresponding structural parameters are plotted against the size of the CIPSI 
expansions. 
We also compare our results with previous QMC calculations~\cite{assaraf2017} of the structural 
optimization of butadiene, which employed various CAS expansions: a CAS(4,4), CAS(4,16), and 
CAS(4,20) correlating 4 $\pi$ electrons in the orbitals constructed from the $2p_z$, $3p_z$, 
$3d_{xz}$, $3d_{yz}$, and $4p_z$ atomic orbitals; a CAS(10,10) consisting of 6 $\sigma$ and 
4 $\pi$ electrons in 10 bonding and antibonding orbitals; a truncated RAS(10,22) that includes 
single and double excitations in additional 12 $\pi$ and $\delta$ orbitals over the CAS(10,10) space.
Since the cc-pVTZ and cc-pVQZ basis sets yield VMC and DMC energies compatible
to better than 2 and 0.2 mHartee, respectively, and structural parameters
differing by less 0.4 m\AA\ (see SI), we can directly compare our cc-pVTZ
calculations with the results obtained with the optimization of Jastrow-CAS
wave functions in the cc-pVQZ basis. 

As shown in Fig.~\ref{fig:energy1}, a CIPSI wave function with roughly 100
determinants yields a VMC energy which is only about 1 and 4 mHartree higher
than the energies of the CAS(10,10) and RAS(10,22) expansions consisting of
15,912 and 45,644 Slater determinants, respectively.  Unlike the Jastrow-CAS
case,
where the energies are scattered around a relatively flat value, CIPSI
expansions of increasing size yield a monotonic decrease in the VMC energy,
with our largest considered expansion of 32,768 determinants amounting to an energy
which is about 40 mHartree lower than the best RAS(10,22) value. In fact, the
VMC energy obtained with our largest CIPSI expansion is within 5 mHartree of
the converged DMC energy obtained with the Jastrow-CAS wave functions.
Therefore, a smart selection of determinants from the approximate FCI space
helps us attain much lower VMC energies 
in comparison to CAS expansions over conventionally used active-space
definitions, which instead lead to the inclusion of many determinants with
little contribution to the energy.

\begin{table*}[t]
\caption{Optimal ground-state structural parameters (\AA) of butadiene and corresponding VMC 
and DMC energies (a.u.) with increasing number of CIPSI determinants obtained in the truncation
scheme. The total number of optimized parameters in the wave function is listed. The statistical error is given in brackets.} 
\label{tab:geom-buta-vmc-dmc-pvtz}
\begin{tabular}{ccccccc}
\hline
          No.\ det  &   No.\ param &     C-C        &      C=C     &      BLA     &     VMC       &  DMC \\  
\hline
               1   &      749        &  1.45595(28)   &  1.32415(11) &  0.13180(30) & -26.24310(32) & -26.30426(28) \\
               2   &      782        &  1.44596(20)   &  1.33025(14) &  0.11571(29) & -26.24912(32) & -26.30681(14) \\
               8   &      822        &  1.45244(28)   &  1.33245(15) &  0.11999(50) & -26.25644(31) & -26.31044(26) \\
             128   &     1594        &  1.45778(12)   &  1.33564(20) &  0.12214(20) & -26.26562(30) & -26.31223(24) \\
            1024   &     5514        &  1.45632(22)   &  1.33493(08) &  0.12139(25) & -26.28829(26) & -26.31908(20) \\
            2048   &     7726        &  1.45626(15)   &  1.33456(09) &  0.12170(17) & -26.29386(25) & -26.32147(16) \\
            5114   &    12147        &  1.45549(08)   &  1.33434(11) &  0.12115(07) & -26.29980(24) & -26.32424(09) \\
           15469   &    24818        &  1.45491(06)   &  1.33406(08) &  0.12085(10) & -26.30880(22) & -26.32873(22) \\ 
           32768   &    44265        &  1.45487(25)   &  1.33414(21) &  0.12072(37) & -26.31194(12) & -26.32928(20) \\
\hline
\end{tabular}
\end{table*}

The behavior of DMC parallels the VMC results with the energy of the Jastrow-CAS
wave functions being lowered by about 8 mHartree when the size of the wave
function is increased from 1 to 20 determinants, and stagnating afterwards as
shown in Figure~\ref{fig:energy1}.  The Jastrow-CIPSI wave function yields a
DMC energy which is comparable with the Jastrow-CAS case when the CIPSI
expansion has only about 100 determinants, while the largest CIPSI expansion
gives a DMC energy 12 mHartree lower than the truncated RAS(10,22) case.  
We also note that the estimate of the FCI
limit in the current basis set on the initial geometry 
is about $-26.275$ Hartree (see SI) and,
therefore, as much as 37 and 54 mHartree higher than our best VMC and DMC
energies. Our best DMC energies are also superior to the CCSD(T) optimized 
values obtained with a   
quintuple-$\zeta$ basis set (see SI).

Importantly, the use of CIPSI expansions (in combination with a Jastrow factor and the
optimization of all wave function parameters) is not only beneficial
in terms of the quality of the final total energies:
the smooth and
monotonic convergence of the VMC and DMC energies with the number
of determinants demonstrates the effectiveness of such
an approach in identifying energetically relevant determinants in a systematic
manner. These important excitations are not easily accessible through manual selection as 
demonstrated by the energy plateau one reaches in constructing very large
expansions based on an {\it a priori} choice of an apparently reasonable set of active 
orbitals.
Besides being completely automated, this feature of the CIPSI scheme is also crucial
for obtaining a smooth convergence of the structural properties (to better than 1 m\AA)
with the number of determinants as discussed next.

In Figure~\ref{fig:geomparams1}, we illustrate the variation of the carbon-carbon
single and double bonds and of the bond length alternation (BLA) obtained with 
the Jastrow-CIPSI expansions. The convergence of the bond lengths and BLA is smooth 
with the increase in the size of the wave function. 
While the double bond is already converged to better than 0.5 m\AA\ with 
little over 2000 determinants, the single bond and, consequently, the BLA show a somewhat 
slower convergence and reach the same level of accuracy with about 5114 
determinants.  We stress however that we are here looking at extremely small
differences while establishing the degree of convergence. For all practical purposes, the
BLA is already converged to better than 1 m\AA\ within the limits of statistical error 
with just over 1000 determinants, again proving the ability of CIPSI expansions to obtain
optimal geometry descriptions with a very small set of determinants. Besides
overcoming the non-uniform convergence of the structural parameters when expanding the 
wave function on different active spaces, these calculations also confirm the
need for a rather subtle multireference description for such conjugated systems~\cite{tenti2017,smith2017}.
%
The converged single bond length is in excellent agreement with the
CCSD(T)/CBS value of 1.455 \AA\ obtained with the same pseudopotentials used in
the QMC calculations (see SI). Our double bond is instead less than 0.003 \AA\ smaller 
than the corresponding CCSD(T)/CBS value, consequently resulting in a difference of 
about 0.002 \AA\ on the BLA. 

To investigate the impact of the inclusion of three-body terms in the Jastrow factor,
we add them to the CIPSI wave functions with 2048 and 5114 determinants
and reoptimize all wave function parameters on the fixed average geometries we have determined 
with the two-body Jastrow factor.  These optimizations result in the VMC and DMC energies presented 
in Table~\ref{tab:3-body-energy}. While there is an expected gain in the VMC energies, the DMC energies 
are equivalent within the statistical error of 0.2 mHartree to the energies obtained with a two-body
Jastrow factor. Consequently, given the quality of our determinantal component, a two-body Jastrow 
recovers most of the missing dynamical correlation contribution and is sufficient for our purposes.

\begin{table}[h!]
\caption{Effect of the inclusion of three-body Jastrow terms on the total VMC and DMC energies (a.u.). 
The structures optimized with the two-body Jastrow factor are used. $\Delta$E denotes the gain in energy with respect to the values obtained with a two-body Jastrow factor.}
\label{tab:3-body-energy}
\begin{tabular}{ccccc}
\hline
          No.\ det & E$_{\rm VMC}$ & E$_{\rm DMC}$ & $\Delta$E$_{\rm VMC}$ & $\Delta$E$_{\rm DMC}$\\  
\hline
            2048   & -26.29908(15) & -26.32162(20) & -0.00522(29) & -0.00015(26)\\
            5114   & -26.30333(23) & -26.32412(16) & -0.00353(33) &\  0.00012(18)\\
\hline
\end{tabular}
\end{table}

\begin{figure}[!htb]
\includegraphics[width=1.0\columnwidth]{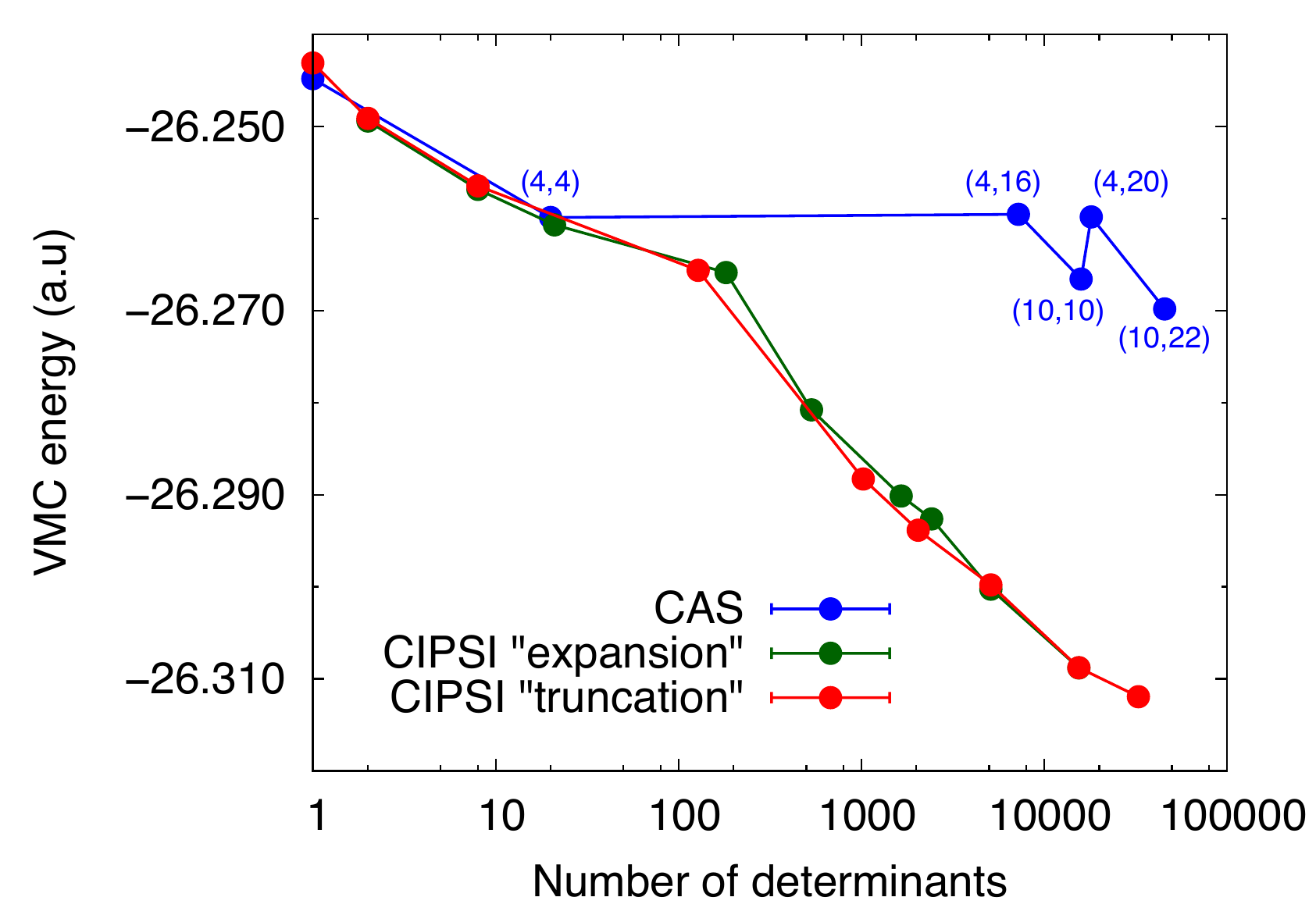}
\caption{Optimal ground-state VMC energies obtained with Jastrow-CAS and Jastrow-CIPSI wave 
functions constructed with the \quotes{expansion} and \quotes{truncation} schemes.}
\label{fig:energyall}
\end{figure}

All results presented so far have been obtained by selecting the CIPSI determinants
out of a much larger CIPSI wave function according to the \quotes{truncation} scheme. 
The VMC energies obtained with the \quotes{expansion} scheme where one constructs 
CIPSI expansions of increasing size are compared with the ``truncation'' scheme in 
Figure~\ref{fig:energyall}. 
Both schemes lead to very comparable convergence in the energy as a function of the 
number of determinants in the Jastrow-CIPSI wave function and to compatible converged
structural parameters. However, as shown in the SI, the variation of the bond lengths 
and BLA is not as smooth 
as in the truncation case.  This difference 
probably arises from the added requirement of having an eigenstate 
of $\textbf{S}^{2}$ in the expansion scheme.  In fact, if we follow the expansion scheme 
without this requirement and construct a wave function of 5114 determinants, we obtain 
an optimal VMC single and double bonds of 1.45564(26) and 1.33496(11) \AA\, respectively,
and, a BLA of 0.12068(26) \AA. The VMC energy converges to -26.29960(24) a.u. 
These results are in good agreement with the corresponding values obtained with the 
truncation scheme.

Finally, we explore a potentially better orbital description in terms of natural orbitals
obtained from a CIPSI calculation instead of the canonical orbitals resulting from a 
CAS(10,10) self-consistent calculation as done above. To this aim, we determine the 
natural orbitals for a very large CIPSI wave function of 2 million determinants and 
construct three expansions of roughly 2000, 5000, and 15,000 determinants 
with the truncation scheme. 
Surprisingly, we consistently obtain higher VMC energies compared
to the expansions generated with canonical orbitals. The reason behind this
observation is unclear. The bond lengths and BLA obtained with 5000
determinants are in excellent agreement with the corresponding values for the
truncated canonical set of comparable size but the other test cases do not provide 
as compatible values. The results of these calculations are given in the SI. 

\section{Conclusion}

We demonstrated the excellent performance of compact perturbatively-selected
CIPSI determinantal expansions in obtaining significantly lower VMC and DMC energies 
as compared to conventional active space definitions for the challenging case of
butadiene. With the use of these wave functions, we were able to obtain converged ground-state structural 
parameters with the use of only a few thousand determinants.  We tested two 
different schemes for the selection of the CIPSI expansions either by constructing 
a large CIPSI wave function and then truncating it (``truncation'' scheme), or by 
considering successive sets of determinants in the CIPSI construction (``expansion'' 
scheme). We found that the two representations are rather equivalent in terms of energy
but that the truncation scheme possibly leads to a somewhat smoother convergence of the
structural parameters with the size of the CIPSI expansion. CCSD(T) calculations with
the same pseudopotentials yield a CBS estimate of the carbon-carbon single bond in very 
good agreement with our converged value but a double bond and corresponding BLA 
smaller by about 0.002 \AA. 
We do not expect any significant change in the bond lengths upon inclusion of additional determinants, in view of their weak variations observed over a wide range of the number of determinants already considered.
 
Our study therefore shows that the automated selection of determinants from a
CIPSI wave function is an extremely suitable and less cumbersome alternative for 
the fast optimization of ground-state geometries in QMC than 
a choice based on correlating electrons in active spaces constructed from energetically 
low-lying orbitals. The latter results in large expansions with many determinants which contribute little
to the energy and to the convergence of the structural parameters. 
The use of CIPSI-based wave functions in combination with the low-scaling
algorithms for simultaneous wave function and geometry optimization opens the way
to the accurate and efficient QMC optimization of large molecular systems.

\section*{Supplementary Information}

See Supplementary Information for CIPSI energies and FCI extrapolation;
VMC and DMC energies of bare CIPSI wave functions; impact of reoptimizing 
the wave function on the average geometry; basis-set convergence of
geometry and corresponding VMC and DMC energies;
bond-length and BLA convergence obtained
with the \quotes{expansion} scheme; energy and geometry convergence with the
use of natural orbitals; DMC time-step extrapolation and CCSD(T)
geometry optimizations with pseudopotential and all-electron basis sets.

\section*{Acknowledgements}

This work is part of the Industrial Partnership Programme (IPP) ``Computational
sciences for energy research'' of the Netherlands Organisation for Scientiﬁc
Research (NWO-I, formerly FOM). This research programme is co-financed by Shell
Global Solutions International B.V. This work was carried out on the Dutch
national supercomputer Cartesius with the support of SURF Cooperative, and
using HPC resources from CALMIP (Toulouse) under allocation 2017-0510.
The authors declare no competing financial interest.

\bibliography{paper_CIPSI}

\begin{thebibliography}{47}
\bibitem [{BFD()}]{BFD_H2013}%
  \BibitemOpen
  \href@noop {} {}\bibinfo {note} {For the hydrogen atom, we use a more
  accurate BFD pseudopotential and basis set. Dolg, M.; Filippi, C., private
  communication}\BibitemShut {NoStop}%
\end{thebibliography}


\end{document}